# Coulomb blockade in a nanoscale phosphorus-in-silicon island

F. E. Hudson [a,*], A. J. Ferguson [a], C. Yang [b], D. N. Jamieson [b], A. S. Dzurak [a], R. G. Clark [a]

[a] *Centre for Quantum Computer Technology, University of New South Wales, Sydney NSW 2052, Australia*

[b] *Centre for Quantum Computer Technology, University of Melbourne, Parkville VIC 3010, Australia*

**Abstract**

We study the low temperature electrical transport behaviour of a silicon single electron transistor. The island and leads are defined by patterned phosphorus doped regions achieved by ion implantation through a polymer resist mask. In the device a 50 nm diameter island, containing ~600 donors and having a metallic density of states, is separated from source and drain leads by undoped silicon tunnel barriers. The central island and tunnel barriers are covered by a surface gate in a field effect transistor geometry allowing the coupling between the leads and island to be controlled. Coulomb blockade due to charging of the doped island is measured, the oscillation period is observed to be constant while the charging energy is dependent on the surface gate voltage. We discuss the possibilities of approaching the few electron regime in these structures, with the aim of observing and manipulating discrete quantum mechanical states.

*Keywords:* coulomb blockade; single electron transistor; quantum computing;

**1. Introduction**

Quantum dots in which few and even single electrons are confined are important systems in which to attempt the control of individual quantum mechanical charge or spin states. To date, GaAs heterostructures, in which the confinement potential is defined using depletion gates on a 2 dimensional electron gas, have provided the model quantum dot system to achieve this coherent control [1-4]. However recent advances in quantum dots fabricated from other materials systems such as Si [5], SiGe [6-8], carbon nanotubes [9,10] and semiconductor nanowires [11,12] also show great promise allow coherent control in the few carrier regime.

Our aim in this work is to approach the few electron regime using low numbers of phosphorus donors in silicon as the confining potential. The devices have a central phosphorus doped island, consisting of ~600 donors, separated by a small gap from ion implanted source-drain leads. Coupling between this central island and the leads is controlled by electrostatic surface gates which are arranged in a field effect transistor (FET) type geometry.

With no potentials applied on the surface gates, there are approximately N = 600 confined electrons with this number reduced slightly due to population of interface traps. Consequently, these dots are in the many electron regime where the observation of single particle states is unlikely. Scaling down to N < 100 electrons in similar devices is possible by reducing the aperture diameter and implant density at which point the effect of single particle states may become important. Using existing fabrication techniques, we have achieved apertures with diameters ~ 10 nm. In addition, using detected single ion implantation we have made considerable progress in fabricating electronic devices that just have single donors, or pairs of donors [13,14]. Measurements on these devices will be described elsewhere.

Previous experiments have used aluminium single electron transistors to investigate the transfer of electrons between islands of donors isolated from electron reservoirs [15] and in transport across larger islands with reservoirs with N ~ 10,000 donors [16]. However this is the first attempt to study transport through ion implanted islands of these dimensions.

The observation of Coulomb blockade in these samples is the main result of this work, and we discuss its parameters and the possibility for measuring islands with fewer phosphorus donor atoms.

## 2. Fabrication

The devices are fabricated from high resistivity silicon ($\rho > 5$ k$\Omega$cm$^{-1}$) with 5 nm of thermally grown SiO$_2$ on the surface. (A completed sample is shown in Figure 1(a).) For this particular oxide grown in our laboratory a trap density of $n_{trap} = 2 \times 10^{11}$ cm$^{-2}$ has been determined [13] by measurement of MOSFET threshold voltages. Before growth of the oxide ohmic contacts are diffused into the silicon, these ensure good electrical contact between the metal bond pads and the ion implanted device leads.

To prepare the samples for ion implantation a polymer implantation mask is defined in 150 nm-thick PMMA resist by electron beam lithography (EBL). The mask contains apertures for a 30 nm diameter dot and source-drain leads separated from the dot by a gap, $d$, that is varied across different devices.

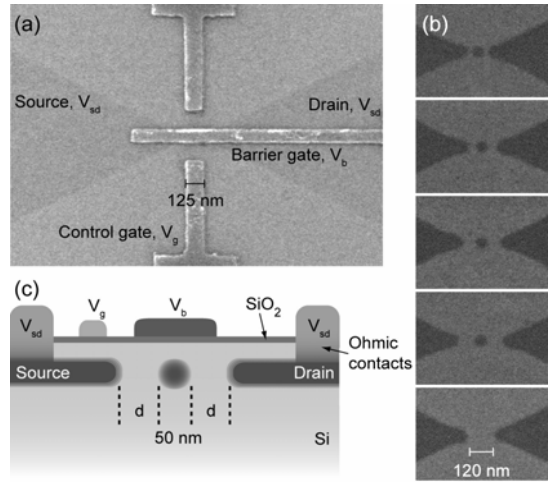

Figure 1: (a) Scanning electron microscope image of a device identical to the ones measured comprising an implanted dot and source-drain leads, barrier control gate, $V_b$, and control gate, $V_g$. (The top-most gate is not used). (b) SEM images of 50 nm implanted dot and leads taken before annealing with gaps from top to bottom of $d$ = 20 ±5 nm, 35 ±5 nm, 55 ±5 nm, 70 ±5 nm and a control device with no dot. (c) Cross-sectional diagram of the device.

Phosphorus is implanted through the mask with an areal dose of $8.5 \times 10^{13}$ cm$^{-2}$ at 14 keV energy resulting in a peak donor concentration at a depth of 20 nm below the substrate surface. The peak expected density of the implanted regions is $n = 4.2 \times 10^{19}$ cm$^{-3}$, which is an order of magnitude greater than the metal-insulator transition for Si:P ($n_{MIT} = 3.45 \times 10^{18}$ cm$^{-3}$). Doping densities greater than the metal insulator transition result in a metallic density of states that allows conduction at low temperature despite negligible thermal excitation of donors.

After removal of the PMMA resist the implanted phosphorus regions are imaged with a scanning electron microscope to record the dimensions of individual devices, as shown in Figure 1(b). In these images the dark areas correspond to a combination of the implanted phosphorus donors and damage caused by the ion implantation process. The images show that the 30 nm diameter apertures yield a damaged region of 50 nm diameter. As a result, the diameter of the implanted islands is taken at 50 nm. The gap between the dot and each lead, $d$, varies between devices by ± 5 nm due to slight differences in the implant mask lithography. A rapid thermal anneal (RTA) performed at 1000 °C for 5s repairs the damage caused by the implantation process while minimising dopant diffusion [18]. The contrast between the implanted region and the undoped silicon decreases significantly after this anneal (seen in Figure 1(a)) consistent with repair of the implantation damage.

In a final EBL step Ti/Au electrostatic gates are defined on the silicon surface. A barrier gate, $V_b$, in a FET geometry, is used to control the turn-on characteristics of the device. Another pair of electrostatic control gates is also added, these also control the potential but are less strongly coupled. By using RTA resistant Ti/Pt markers an alignment precision of less than 50 nm between the ion implanted region and the surface gates is routinely achieved.

## 3. Measurement

The electrical transport measurements through the device were performed at the base temperature of a dilution refrigerator (T ~ 100 mK) using standard low-frequency (f < 100Hz) lock-in techniques with an ac excitation voltage $V_{ac}$ = 20 μV.

In order to characterise the coupling of the implanted island to the leads, initial measurements were made to determine the device conductance as a function of the barrier gate voltage. This is identical to measuring the turn-on characteristic of a MOSFET.

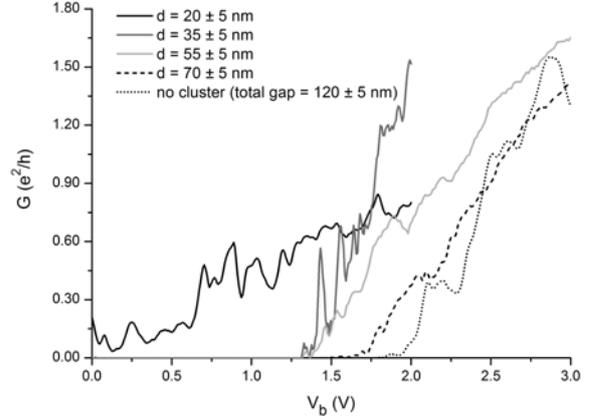

Figure 2: Conductance through source-drain leads as a function of barrier gate voltage for devices with different values of the dot to lead gap, $d$. There is a general trend for turn-on voltage to increase with $d$. Devices with $d$ = 20 ± 5 nm are conducting at $V_b$ = 0 V

The turn-on curve for a device with no island is shown in Figure 2. This device constitutes a nanoscale MOSFET and has a gap between the source and drain electrodes of 120 nm with threshold occurring at $V_b$ = 1.9 V. However, as seen by other investigators [19,20], for MOSFETs of these dimensions there are a number of oscillations in the turn-on curve. This has been attributed to Coulomb blockade of the random disorder potential in the device channel, and can be viewed as the local potential minima in the channel undergoing single electron charging. As a result of the disordered nature of the interface, the potential landscape changes significantly with barrier gate voltage leading to aperiodic Coulomb blockade with a variable charging energy.

Figure 2 also compares this behaviour of this control sample to devices that include the ion implanted island between the source-drain leads. The turn-on curves shown here are for devices with a different spacing, $d$, between the dot and the leads. These measurements are qualitatively the same as the control sample: there is an increasing conductance with barrier gate voltage and in all cases there is evidence of Coulomb blockade in the turn-on characteristics.

A trend of decreasing threshold voltage is seen as the separation of the island to the leads is reduced. For the smallest separation, $d = 20 \pm 5$ nm, there is measurable conductance at $V_g = 0$ V and a negative voltage is required to turn the channel off. While there is clear evidence for a trend in the turn-on voltage, there is also significant variation in this threshold between devices with nominally identical spacings between the leads and the island.

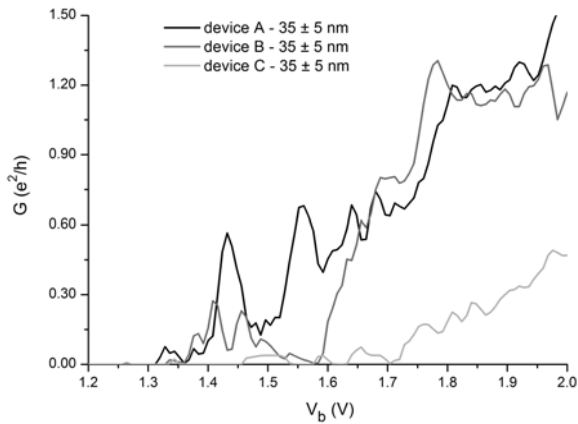

Figure 3: Variation in turn-on voltage characteristics from three devices with the same nominal gap, $d = 35 \pm 5$ nm.

Three traces for different devices that were all fabricated to have 35 nm spacing between the leads and island are shown in Figure 3. For these samples there is a spread in threshold voltages of 0.4 V, with at least part of this variation likely to be due to the same random potential that causes the Coulomb blockade, and part being due to variability in the processing.

In order to investigate Coulomb blockade in the phosphorus doped islands we set the barrier gate $V_b$ near to the threshold voltage and sweep its value over a small range (see Figure 4). This device is the same one as device C in Figure 3. In doing this we find almost periodic oscillations in the conductance with a period of $\Delta V_b = 3.1 \pm 0.3$ mV, corresponding to a capacitance $C_b = 52$ aF.

These oscillations can be distinguished from Coulomb blockade due to the disorder potential by their periodicity. In addition when similar sweeps are performed at different values of $V_b$ oscillations with this same period are found. So while there are clearly other blockade effects occurring in the undoped tunnel barriers between the leads and the island, periodic charging of the central island is a persistent effect. The peak conductance of these Coulomb blockade oscillations is not constant which is probably indicative of the other blockade effects occurring in the undoped tunnel barriers. In control devices with no island in the channel, this nearly periodic charging with a few mV period was not observed.

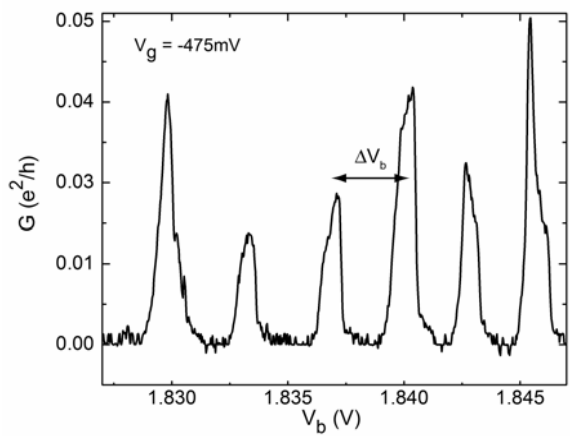

Figure 4: Single trace over Coulomb blockade peaks at zero source-drain bias for a device with $35 \pm 5$ nm gaps showing a regular period $\Delta V_b = 3.1 \pm 0.3$ mV.

These oscillations are also affected by the voltage, $V_g$, on the control gate showing a period $\Delta V_g = 115$ mV, and a much smaller capacitance $C_g = 1.4$ aF. Adjusting the potential on this gate has the additional effect of slightly modifying the channel properties. For example, we find that the behaviour of the device is less noisy at certain voltages on this gate, for this reason the control gate voltage is set to $V_g = -475$ mV for these measurements.

By measuring the differential conductance as a function of source-drain bias across the island we can map out the Coulomb diamonds (Figure 5). The diamonds have a charging energy, $E_c = e^2/2C_\Sigma = 415 \pm 66$ μeV. It is interesting to note that as $V_b$ is changed the oscillation period remains constant, however a considerable change is noticed in the charging energy. At $V_b$ = 2.3 V a significantly reduced value of $E_c$ = 32 ± 28 μeV is measured, indicating an order of magnitude increase in the total capacitance. This can be understood by the electron gas in the MOSFET channel surrounding the island more completely at higher gate biases and therefore increasing the total capacitance while not significantly altering the gate capacitance.

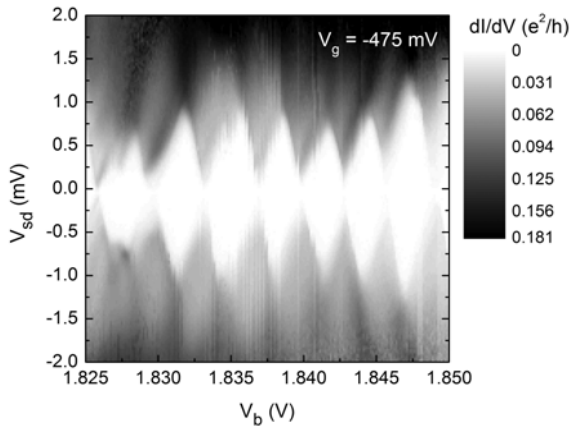

Figure 5: Coulomb diamonds for the same device is in Figure 4. Charging energy of the island in this voltage range is, $E_c$ = 415 ± 66 μeV.

Depending on the barrier gate voltage, the number of electrons in the island may differ greatly from the number of dopants that define the confinement potential. For a constant Coulomb oscillation period of 3.1 mV, and assuming that at $V_g$ = 0 V there are 600 confined electrons, at $V_g$ = 1.875 V there could be an approximate total of 1200 confined electrons. This indicates that the devices with smaller separations, $d$ = 20 ± 5 nm, are likely to be better candidates for few electron quantum dots measured by electrical transport. In this case it is possible to obtain measurable values of conductance without setting a large gate voltage and consequently adding a large number of electrons to the system. In the $d$ = 20 ± 5 nm device, we measured regular Coulomb oscillations with a period of $\Delta V_b$ = 6.8 ± 0.4 mV.

## 4. Conclusions

Single electron charging has been observed by transport measurements on phosphorus in silicon islands containing 600 ion implanted donors. The experiments found Coulomb blockade due to charging of the island. This could be distinguished from charging of the disordered interface potential by its nearly periodic behaviour. We note that in control devices without the island, only aperiodic charging could be observed.

Our future research will attempt to scale the islands towards the few electron level where discrete quantum levels may be observed. In order to achieve this, our starting point is to reduce the number of ion implanted donors. Whilst still maintaining a metallic density of states on the island and using the same aperture diameter, the ion implantation dose can be reduced to create a dot of 175 donors through which transport can be studied. In addition, reduction of the aperture diameter to $d$ = 10 nm yields dots defined by just 20 donors.

### Acknowledgements

The authors would like to thank S. E. Andresen, V. Chan and R. Brenner for helpful discussions and E. Gauja, R. P. Starrett, D. Barber, G. Tamanyan and R. Szymanski for their technical support. This work is supported by the Australian Research Council, the Australian Government and by the US National Security Agency (NSA), Advanced Research and Development Activity (ARDA) and the Army Research Office (ARO) under contract number DAAD19-01-1-0653.